\documentclass[10pt]{article}
\usepackage{amsmath,amssymb}
\usepackage{url}
\usepackage{graphicx}
\usepackage{subfigure}
\usepackage{cite}
\usepackage{color}
\usepackage{comment}
\usepackage[toc,page]{appendix}

\begin{document}

\title{Methods of Information Theory and Algorithmic Complexity for Network Biology\thanks{Dr. Ali Masoudi-Nejad, sole Guest Editor for this paper. Forthcoming in the journal of \textit{Seminars for Developmental Cell Biology} special issue on Information Theory in Molecular Biology.}}

\author{Hector Zenil\footnote{Corresponding author: hector [dot] zenil [at] algorithmicnaturelab [dot] org}, Narsis A. Kiani and Jesper Tegn\'er\\
Unit of Computational Medicine, Department of Medicine,\\ Karolinska Institute \&
Center for Molecular Medicine,\\Karolinska University Hospital, Stockholm, Sweden.}

\date{}

\maketitle


\begin{abstract}
We survey and introduce concepts and tools located at the intersection of information theory and network biology. We show that Shannon's information entropy, compressibility and algorithmic complexity quantify different local and global aspects of synthetic and biological data. We show examples such as the emergence of \textit{giant components} in Erd{\"o}s-R\'enyi random graphs, and the recovery of topological properties from numerical kinetic properties simulating gene expression data. We provide exact theoretical calculations, numerical approximations and error estimations of entropy, algorithmic probability and Kolmogorov complexity for different types of graphs, characterizing their variant and invariant properties. We introduce formal definitions of complexity for both labeled and unlabeled graphs and prove that the Kolmogorov complexity of a labeled graph is a good approximation of its unlabeled Kolmogorov complexity and thus a robust definition of graph complexity.\\

\end{abstract}

\section{Introduction}


Over the last decade network theory has become a unifying language in biology, giving rise to whole new areas of research in computational systems biology. Gene networks are conceptual models of genetic regulation where each gene is considered to be directly affected by a number of other genes, and are usually represented by directed graphs. 


Classical information theory has for some time been applied to networks, but Shannon entropy, like any other computable measure (i.e. one that is a total function, returning an output in finite time for every input), is not invariant to changes of object description~\cite{smalldata}. 

More recently, algorithmic information theory has been introduced as a tool for use in network theory, and some interesting properties have been found~\cite{zenilgraph,zenilkiani,journalcomplexnetworks}. For example, in \cite{zenilgraph} correlations were reported among algebraic and topological properties of synthetic and biological networks by means of algorithmic complexity, and an application to classify networks by type was developed in \cite{zenilkiani}. 

We review and explore further these information content approaches for characterizing biological networks and networks in general. We provide theoretical estimations of the error of approximations to the Kolmogorov complexity of graphs and complex networks, offering both exact and numerical approximations. Together with~\cite{zenilgraph} and~\cite{zenilkiani}, the methods introduced here represent a novel view and constitute a formal approach to graph complexity, while providing a new set of tools for the analysis of the local and global structure of networks.


\subsection{Graph notation and complex networks}

\begin{figure}[htbp!]
\centering
\includegraphics[width=5.6cm]{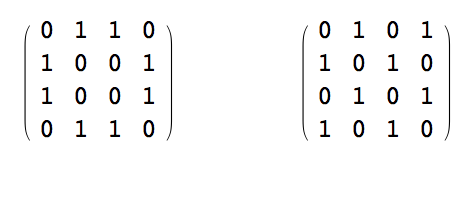}\\

\includegraphics[width=5.8cm]{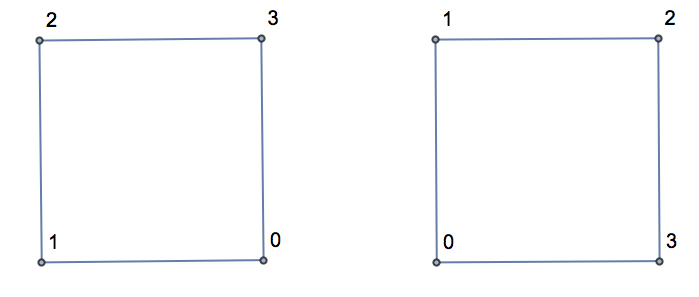}
\caption{Isomorphic graphs with two different adjacency matrix representations, illustrating that the adjacency matrix is not an invariant of a graph under relabelings. However, similar graphs have adjacency matrices with similar algorithmic information content.}
\label{isotests}
\end{figure}

A graph $G$ is \textit{labeled} when the vertices are distinguished by names such as $u_1, u_2, \ldots u_n$ with $n=|V(G)|$. Graphs $G$ and $H$ are said to be \textit{isomorphic} if there is a bijection between the vertex sets of $G$ and $H$, $\lambda : V(G) \rightarrow V(H)$ such that any two vertices $u$ and $v \in G$ are adjacent in $G$ if and only if $\lambda(u)$ and $\lambda(v)$ are adjacent in $H$. When $G$ and $H$ are the same graph, the bijection is referred to as an \textit{automorphism} of $G$. The adjacency matrix of a graph is not an invariant under \textit{graph relabelings}. Fig.~\ref{isotests} shows two adjacency matrices for isomorphic graphs. A \textit{canonical form} of $G$ is a labeled graph $Canon(G)$ that is isomorphic to $G$, such that every graph that is isomorphic to $G$ has the same canonical form as $G$. An advantage of $Canon(G)$ is that unlike $A(G)$, $A(Canon(G))$ is a graph invariant of $Canon(G)$~\cite{canon}.

One of the most basic properties of graphs is the number of links per node. When all nodes have the same number of links, the graph is said to be \textit{regular}. The \textit{degree} of a node $v$, denoted by $d(v)$, is the number of (incoming and outgoing) links to other nodes. We will also say that a graph is \textit{planar} if it can be drawn in a plane without its edges crossing. Planarity is an interesting property because only planar graphs have \textit{duals}. A \emph{dual graph} of a planar graph $G$ is a graph that has a vertex corresponding to each face of $G$, and an edge joining two neighboring faces for each edge in $G$.

\begin{figure}[htbp!]
\centering
\includegraphics[width=3.3cm]{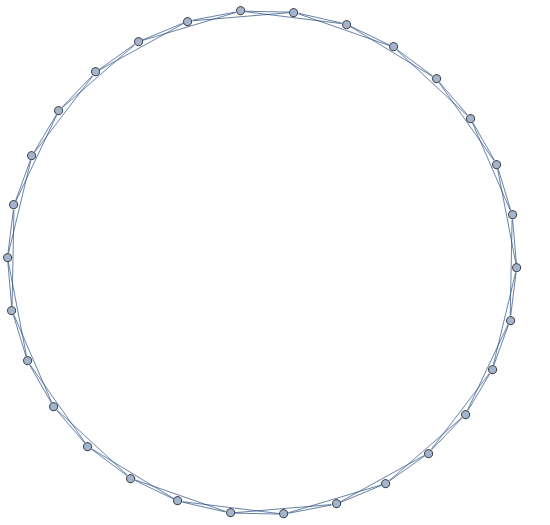}
\includegraphics[width=3.8cm]{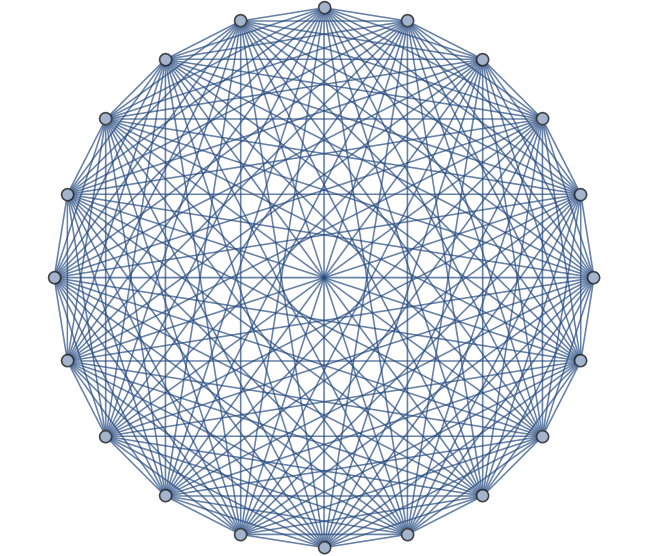}
\includegraphics[width=4.2cm]{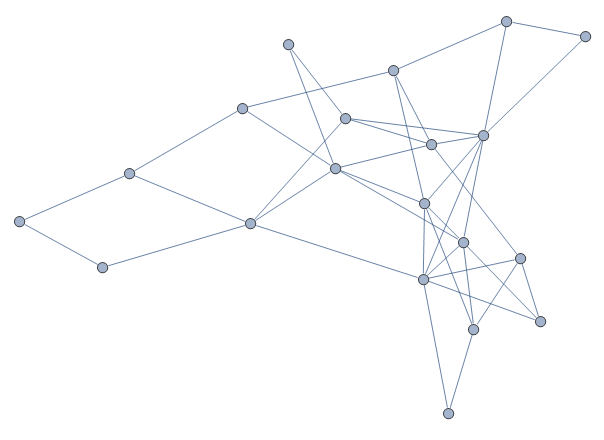}
\caption{Examples of two regular graphs (left and middle) are a $2n$ circular graph with 20 nodes and a complete graph with 20 nodes, both of whose descriptions are very short, hence $K(G) \sim \log |V(G)| \sim 4.32$ bits. In contrast, a random graph (right) with the same number of nodes and number of links requires more information to be specified, because there is no simple rule connecting the nodes and therefore $K(G) \sim |E(G)| = 30$ bits.}
\label{fig:regularrandom}
\end{figure}

A popular type of graph that has been studied is the so-called \textit{Erd{\"o}s-R\'enyi}~\cite{erdos,gilbert} ($ER$) graph, in which vertices are randomly and independently connected by links with a fixed probability (also called \textit{edge density}) (see Fig.~\ref{fig:regularrandom} for a comparison between a regular and a random graph of the same size). The probability of vertices being connected is called the \textit{edge probability}. The main characteristic of random graphs is that all nodes have roughly the same number of links, equal to the average number of links per node. A $ER$ graph $G(n, p)$ is a graph of size $n$ constructed by connecting nodes randomly with probability $p$ independent from every other edge. Usually $ER$ graphs are assumed to be non-recursive (i.e. truly random), but $ER$ graphs can be constructed recursively with, for example, pseudo-random algorithms. Here we will assume that $ER$ graphs are non-recursive, as theoretical comparisons and bounds hold only in the non-recursive case. For numerical estimations, however, we use a pseudo-random edge connection algorithm, in keeping with common practice.

$ER$ random graphs have some interesting properties, but biological networks are not random. They carry information, connections between certain elements in a biological graph are favored or avoided, and not all vertices have the same probability of being connected to other vertices. The two most popular complex network models consist of two algorithms that reproduce certain characteristics found in empirical networks. Indeed, the field has been driven largely by the observation of properties that depart from properties modeled by regular and random graphs. Specifically, there are two topological properties of many complex networks that have been a focus of interest. A \textit{simple graph} is a graph with no self-loops and no multi-edges. Throughout this paper we will only consider simple graphs.

\begin{figure}[htbp!]
\centering
\includegraphics[width=11.5cm]{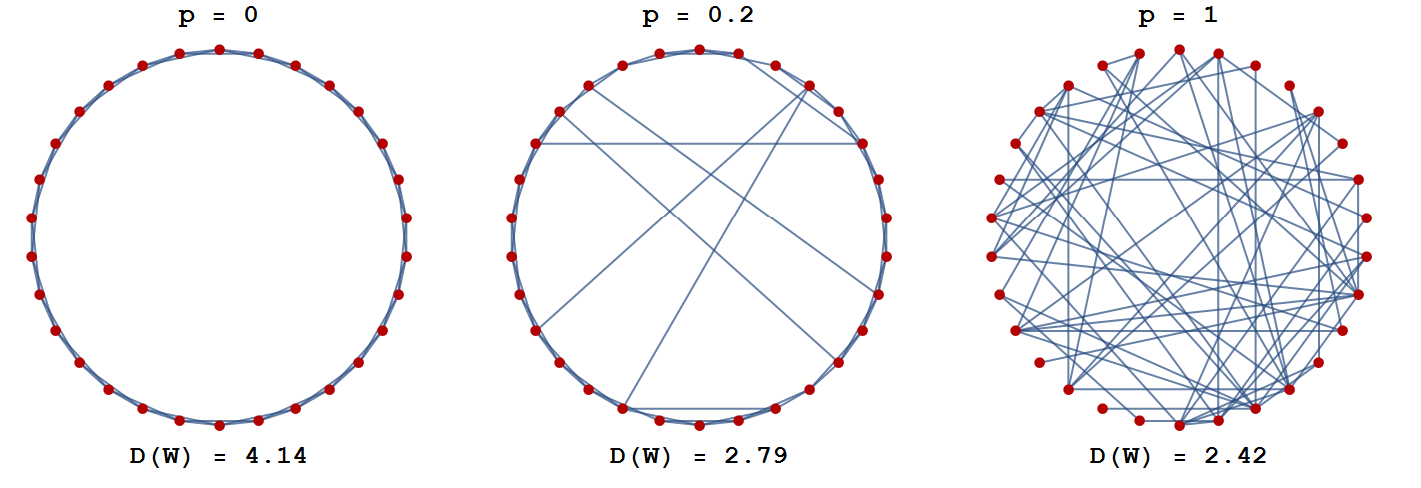}
\caption{For $p=0$, $WS$ is a regular graph that requires little information to be described. For rewiring probability $p=1$, $WS$ becomes a random graph that requires about as much information in bits as the number of edges in $WS$ (i.e. $K(WS) \sim |E(WS)|$). For $p$ very close to zero, however, the average network distance among nodes drops dramatically, approaching that of the random graph case while remaining of low complexity (little information is required to describe $K(WS) \sim \log |V(WS)| + \epsilon(|V(WS)|)$), close to the regular $WS$ for $p=0$.}
\label{wsplot}
\end{figure}

A network is considered a \textit{small-world} graph $G$ (e.g. see Fig.~\ref{wsplot}) if the average graph distance $D$ grows no faster than the $\log$ of the number of nodes: $D \sim \log V(G)$. Many networks are \textit{scale-free}, meaning that their degrees are size independent, in the sense that the empirical degree distribution is independent of the size of the graph up to a logarithmic term. That is, the proportion of vertices with degree $k$ is proportional to $\gamma k^\tau$ for some $\tau > 1$ and constant $\gamma$. In other words, many empirical networks display a power-law degree distribution.

\section{Classical information theory and linear complexity}
\label{classical}

Central to information theory is the concept of Shannon's information entropy, which quantifies the average number of bits needed to store or communicate a message. Shannon's entropy determines that one cannot store (and therefore communicate) a symbol with $n$ different symbols in less than $\log(n)$ bits. In this sense, Shannon's entropy determines a lower limit below which no message can be further compressed, not even in principle. Another application (or interpretation) of Shannon's information theory is as a measure for quantifying the \emph{uncertainty} involved in predicting the value of a random variable. For example, specifying the outcome of a fair coin flip (two equally likely outcomes) requires one bit at a time, because the results are independent and therefore each result conveys maximum entropy. Things begin to get interesting when the coin is not fair. If one considers a coin with heads on both obverse and reverse, then the tossing experiment always results in heads, and the message will always be 1 with full ceryainty.

For an ensemble $X(R, p(x_i))$, where $R$ is the set of possible outcomes (the random variable), $n=|R|$ and $p(x_i)$ is the probability of an outcome in $R$. The Shannon information content or entropy of $X$ is then given by

$$H(X)=-\sum_{i=1}^n p(x_i) \log_2 p(x_i) \textnormal{ (Eq. 1)}$$

Which implies that to calculate $H(X)$ one has to know or assume the mass distribution probability of ensemble $X$. One caveat regarding Shannon's entropy is that one is forced to make an arbitrary choice regarding granularity. Take for example the bit string 01010101010101. The Shannon entropy of the string at the level of single bits is maximal, as there are the same number of 1s and 0s, but the string is clearly regular when two-bit blocks are taken as basic units, in which instance the string has minimal complexity because it contains only 1 symbol (01) from among the 4 possible ones (00,01,10,11). One way to overcome this problem is to take into consideration all possible ``granularities'' (we call this \textit{Block entropy}), from length 1 to $n$, where $n$ is the length of the sequence. This measure is related to what's also called \textit{predictive information} or \textit{excess entropy} (the differences among the entropies for consecutive block sizes). To proceed by means of block entropy is computationally expensive, as compared to fixing the block size to $n$, as it entails producing all possible overlapping ${{i}\choose{n}}$ substrings for all $i \in \{1,\ldots,n\}$.

Shannon entropy can also be applied to the node degree sequence of a graph. The Shannon entropy of an unlabeled network as described by its degree distribution can be described by the same formula for Shannon entropy where the random variable is a degree distribution. The chief advantage of so doing being that it is invariant to relabelings. This also means that the degree distribution is not a lossless representation of a labeled network (but of its isomorphic group), and is an interesting entropic measure but can only be used when the node labels are not relevant. For example, in a causal recursive network, the node labels may represent time events in a sequence that has a meaning captured in the network labeling, in which case the degree distribution sequence (where no agreement has been reached on the order of the elements in the distribution sequence, which is therefore disordered) cannot be used to reconstruct the original network represented by the unlabeled version or the isomorphism group of the labeled networks.

It is also clear that the concept of entropy rate cannot be applied to the degree distribution, because the node degree sequence has no particular order or any order is meaningless because any label numbering will be arbitrary. This also means that Shannon entropy is not invariant to the description of a network, especially as a labeled or an unlabeled network, except for clearly extreme cases (e.g. fully disconnected and completely connected networks both have flat degree distributions and therefore lowest Shannon entropy both degree sequence and adjacency matrix density).

A review and comparison of these entropic measures is provided in~\cite{dehmer} in connection to applications of drug and chemical structures. 

An interesting proposal related to algorithmic complexity is a measure of linear matrix complexity~\cite{neel} based on the number of operations needed to produce a matrix. This is similar to finding the generating basis (a lossless description of a labelled graph) of an adjacency matrix $A$, such that $X M = A$ with $X$ a proper vector such that $X M \leq A$. $X M \leq A$ is an upper bound to the Kolmogorov complexity if a labelled graph $G$ with adjacency matrix $A$.

Effectively, the null space, kernel or vector space basis that generates the adjacency matrix is a smaller computer program that generates the adjacency matrix. This kind of complexity is linear because the matrix could be reduced to non-linear combinations. Indeed, the nullspace is not the shortest possible program to produce $A$.

\section{Algorithmic information and network biology}
\label{kolmo}

DNA sequences store the information required to produce the proteins needed by living organisms, among other information regulating underlying mechanisms, and are therefore expected to be removed from randomness.

Likewise, biological networks carry information, transfer information from one region to another, and implement functions represented by the element's interactions. Connections among elements in a biological network are therefore removed both from triviality and randomness. In a biological network, nodes usually represent proteins, metabolites, genes, transcription factors, etc. A link represents the interactions between the nodes in a network that can correspond to protein-protein binding interactions, metabolic coupling or regulation.

The Shannon entropy (or simply entropy) of a graph $G$ is simply defined by 

$$H(A(G))=-\sum_{i=1}^n P(A(x_i)) \log_2 P(A(x_i)) \textnormal{ (Eq. 2)}$$

\noindent where $G$ is the random variable with $n$ possible outcomes (all possible adjacency matrices of size $|V(G)|$). For example, a completely disconnected graph $G$ with all adjacency matrix entries equal to zero has entropy $H(A(G))=0$, because the number of different symbols in the adjacency matrix is 1. However, if a different number of 1s and 0s occur in $A(G)$, then $H(A(G))\neq0$. In general we will use Block entropy in order to detect more graph regularities (through the adjacency matrix) at a greater resolution. But for Block entropy there is an extra factor to be taken into account. The unlabeled calculation of the Block entropy (not relevant for 1-bit entropy) of a graph has to take into consideration all possible adjacency matrix representations for all possible labelings. Therefore, the Block entropy of a graph is given by:

$$
H(G)=\min\{H(A(g_L)) | G_L \in L(G)\}
\textnormal{ (Eq. 3)}$$

\noindent where $L(G)$ is the group of all possible labelings of $G$. 

Likewise, the degree of a node in a biological network of protein interactions represents the number of proteins with which it interacts, and its Shannon entropy does not require a labeling version because the degree sequence is an invariant of the graph isomorphic group.

Other entropic measures of network elements are possible and have been proposed before but they all require to focus on a particular graph element (adjacency matrix, degree sequence, number of bifurcations) and they do not all converge meaning that Shannon entropy is not invariant to different descriptions of the same object.

Another powerful measure of information content and randomness, with which to find not only statistical but recursive regularities, is the concept of Kolmogorov complexity~\cite{kolmo,chaitin}---denoted by $K$. This is because $K$ has been proven to be a universal measure theoretically guaranteed to asymptotically identify any \textit{computable regularity}~\cite{solomonoff}, i.e. a regularity that can be reproduced by algorithmic means such as a computer program or Turing machine. Formally, the Kolmogorov complexity of a string $s$ is 

$$K(s)=\min\{|p| : U(p)=s\} \textnormal{ (Eq. 4)}$$

\noindent that is, the length (in bits) of the shortest program $p$ that when running on a universal Turing machine $U$ outputs $s$ upon halting.

A universal Turing machine $U$ is an abstraction of a general-purpose computer that can be programmed to reproduce any computable object, such as a string or a network (e.g. the elements of an adjacency matrix). By the \emph{invariance theorem}~\cite{calude,li}, $K_U$ only depends on $U$ up to a constant, so as is conventional, the $U$ subscript can be dropped. Formally, $\exists \gamma$ such that $|K_U(s) - K_{U\prime}(s) | < \gamma$ where $\gamma$ is a constant independent of $U$ and $U\prime$. Because everything in the theory of Kolmogorov complexity is meant to be asymptotic, this invariance theorem means that the longer the string $s$ the closer the Kolmogorov complexity evaluations, even for different Turing machines $U$ and $U\prime$, and at the limit (for $|s| \rightarrow \infty$) the evaluations will coincide.

Due to its power, $K$ comes burdened with a technical inconvenience (formally called \textit{semi-computability}) and it has been proven that no effective algorithm exists which takes a string $s$ as input and produces the exact integer $K(s)$ as output~\cite{kolmo,chaitin}. This is related to a common problem in computer science known as the undecidability of the halting problem~\cite{turing}---referring to the ability to know whether or not a computation will eventually stop.

Despite the inconvenience, $K$ can be effectively approximated by using, for example, compression algorithms. Kolmogorov complexity can alternatively be understood in terms of uncompressibility. If an object, such as a biological network, is highly compressible, then $K$ is small and the object is said to be non-random. However, if the object is uncompressible then it is considered algorithmically random.

A compression ratio, also related to the \textit{randomness deficiency} of a network or how removed an object is from maximum algorithmic randomness, will be defined by $C(G)=Comp(G)/|A(G)|$, where $Comp(G)$ is the compressed length in bits of the adjacency matrix $G$ of a network using a \emph{lossless} compression algorithm (e.g. Compress), and $|A(G)|$ is the size of the adjacency matrix measured by taking the dimensions of the array and multiplying its values. e.g., if the adjacency matrix is 10 $\times$ 10, then $|A(G)|=100$. It is worth mentioning that compressibility is a sufficient test for non-randomness. Which means that it does not matter which lossless compression algorithm is used, if an object can be compressed then it is a valid upper bound of its Kolmogorov complexity. Which in turn means that the choice of lossless compression algorithm is not very important, because one can test them one by one and always retain the best compression as an approximation to $K$. A lossless compression algorithm is an algorithm that includes a decompression algorithm that retrieves the exact original object, without any loss of information when decompressed. The closer $C(G)$ is to 1 the less compressible, and the closer to 0 the more compressible.

In Table~\ref{table}, Kolmogorov approximations by compression of 22 metabolic networks from~\cite{barabasinets}, including the Streptococcus pyogenes network, have been calculated.

\subsection{Algorithmic probability}
\label{bdm}

Another seminal concept in the theory of algorithmic information, is the concept of \textit{algorithmic probability}~\cite{solomonoff,levin} and its related so-called \textit{Universal distribution}~\cite{kirchherr}, also known as ``Levin's \textit{probability semi-measure}''~\cite{levin}.

The algorithmic probability of a string $s$ provides the probability that a valid random program $p$ written in bits uniformly distributed produces the string $s$ when run on a universal (prefix-free~\footnote{The group of valid programs forms a prefix-free set (no element is a prefix of any other, a property necessary to keep $0 < m(s) < 1$.) For details see~\cite{cover,calude}.}) Turing machine $U$. Formally,

$$m(s) = \sum_{p:U(p) = s} 1/2^{|p|} \textnormal{ (Eq. 5)}$$

That is, the sum over all the programs $p$ for which $U$ outputs $s$ and halts.

The algorithmic probability measure $m(s)$ is related to Kolmogorov complexity $K(s)$ in that $m(s)$ is at least the maximum term in the summation of programs, given that the shortest program carries the greatest weight in the sum. The algorithmic Coding Theorem~\cite{levin} further establishes the connection between $m(s)$ and $K(s)$ as follows: 

$$|-\log_2 m(s) - K(s)| < \mathcal{O}(1) \textnormal{ (Eq. 6)}$$

\noindent where $\mathcal{O}(1)$ is an additive value independent of $s$. The Coding Theorem implies that~\cite{cover,calude} one can estimate the Kolmogorov complexity of a string from its frequency. By rewriting Eq.~(5) as: 

$$K_m(s)=-\log_2 m(s) + \mathcal{O}(1) \textnormal{ (Eq. 7)}$$

\noindent one can see that it is possible to approximate $K$ by approximating $m$ (which is why it is denoted $K_m$), with the added advantage that $m(s)$ is more sensitive to small objects~\cite{d4} than the traditional approach to $K$ using lossless compression algorithms, which typically perform poorly for small objects (e.g. small graphs).

\subsection{Kolmogorov complexity of unlabeled graphs}

As shown in~\cite{zenilgraph}, estimations of Kolmogorov complexity are able to distinguish complex from random networks (of the same size, or growing asymptotically), which are both in turn distinguished from regular graphs (also of the same size). $K$ calculated by the BDM assigns low Kolmogorov complexity to regular graphs, medium complexity to complex networks following Watts-Strogatz or Barab\'asi-Albert algorithms, and higher Kolmogorov complexity to random networks. That random graphs are the most algorithmically complex is clear from a theoretical point of view: nearly all long binary strings are algorithmically random, and so nearly all random unlabeled graphs are algorithmically random~\cite{randomgraphs}, where Kolmogorov complexity is used to give a proof of the number of unlabeled graphs as a function of its randomness deficiency (how far it is from the maximum value of $K(G)$).

The \textit{Coding Theorem Method} (CTM)~\cite{d4,d5} is rooted in the relation provided by algorithmic probability between frequency of production of a string from a random program and its Kolmogorov complexity (Eq.~(6). It is also called the algorithmic \textit{Coding theorem}, to contrast it with another coding theorem in classical information theory). Essentially it uses the fact that the more frequent a string (or object), the lower its Kolmogorov complexity; and strings of lower frequency have higher Kolmogorov complexity.

In~\cite{zenilgraph}, numerical evidence was provided in support of the theoretical assumption that the Kolmogorov complexity of an unlabeled graph should not be far removed from that of any of its labeled versions. This is because there is a small computer program of fixed size that should determine the order of the labeling proportional to the size of the isomorphism group. Indeed, when the isomorphism group is large, the labeled networks have more equivalent descriptions given by the symmetries, and would therefore, according to algorithmic probability, be of lower Kolmogorov complexity.

\subsection{Reconstructing $K$ from local graph algorithmic patterns}

The approach to determining the algorithmic complexity of a graph thus involves considering how often the adjacency matrix of a motif is generated by a random Turing machine on a 2-dimensional array, also called a \textit{termite} or \textit{Langton's ant}~\cite{langton}. We call this the \textit{Block Decomposition Method} (BDM) as it requires the partition of the adjacency matrix of a graph into smaller matrices using which we can numerically calculate its algorithmic probability by running a large set of small 2-dimensional deterministic Turing machines, and then, by applying the algorithmic Coding theorem, its Kolmogorov complexity. Then the overall complexity of the original adjacency matrix is the sum of the complexity of its parts, albeit with a logarithmic penalization for repetitions, given that $n$ repetitions of the same object only adds $\log n$ to its overall complexity, as one can simply describe a repetition in terms of the multiplicity of the first occurrence. More formally, the Kolmogorov complexity of a labeled graph $G$ by means of $BDM$ is defined as follows:

\begin{equation}
\label{newecaeq}
K_{BDM} (G,d) = \sum_{(r_u,n_u)\in A(G)_{d\times d}} \log_2(n_u)+K_m(r_u)
\end{equation}
where $K_m(r_u)$ is the approximation of the Kolmogorov complexity of the subarrays $r_u$ arrived at by using the algorithmic Coding theorem (Eq.~(6)), while $A(G)_{d\times d}$ represents the set with elements $(r_u,n_u)$, obtained by decomposing the adjacency matrix of $G$ into non-overlapping squares of size $d$ by $d$. In each $(r_u,n_u)$ pair, $r_u$ is one such square and $n_u$ its multiplicity (number of occurrences). From now on $K_{BDM} (g,d=4)$ will be denoted only by $K(G)$, but it should be taken as an approximation to $K(G)$ unless otherwise stated (e.g. when taking the theoretical true $K(G)$ value). Once CTM is calculated, BDM can be implemented as a lookup table, and hence runs efficiently in linear time for non-overlapping fixed size submatrices.

As with Block entropy (c.f. Section~\ref{kolmo}), the Kolmogorov complexity of a graph $G$ is given by:

$$K^\prime(G)=\min\{K(A(G_L)) | G_L \in L(G)\}
\textnormal{ (Eq. 8)}$$

\noindent where $L(G)$ is the group of all possible labelings of $G$ and $G_L$ a particular labeling. In fact $K(G)$ provides a choice for graph canonization, taking the adjacency matrix of $G$ with the lowest Kolmogorov complexity. Because it may not be unique, it can be combined with the smallest lexicographical representation when the adjacency matrix is concatenated by rows, as is traditionally done for graph canonization. 

By taking subarrays of the adjacency matrix we ensure that network motifs (overrepresented graphs), used in biology and proven to classify superfamilies of networks~\cite{alonreview,milo}, are taken into consideration in the BDM calculation. And indeed, we were able to show that BDM alone classifies the same superfamilies of networks~\cite{zenilkiani} that classical network motifs were able to identify.

Of course the time complexity of the calculation of $K^\prime(G)$ is a matter of concern. The \textit{graph isomorphism problem} involves constructing an `efficient' (subexponential) algorithm for testing whether two given graphs are isomorphic. The graph isomorphism problem is believed to be in \textbf{NP}. The problem of testing labeled graphs for isomorphism is polynomially equivalent to the graph isomorphism problem, and therefore the calculation of $K^\prime(G)$ cannot be expected to have a polynomial time algorithm. We will see, however, that $|K(G)-K^\prime(G)|$ is bounded by a constant independent of $G$.

\section{Small patterns in biological networks}
\label{motifs}

One important development in network biology is the concept of network motifs~\cite{alon,zenildemo}, defined as recurrent and statistically significant sub-graphs found in networks, as compared to a uniform distribution in a random network. As is to be expected, biological networks are not random networks, because biological networks carry information necessary for an organism to develop. Motifs are believed to be of signal importance largely because they may reflect functional properties. 


For example, the function of the so-called FFL (\textit{Feed-Forward Loop}) has been speculated to be a more stable transmitter of information among genes than other possible small network arrangements~\cite{alonreview}. This motif consists of three genes: a regulator that regulates another regulator and a gene, this latter regulated by the two regulators (given that each of the regulatory interactions can be either an activation or a repression, this motif can be divided into 8 more refined subtypes). This 3-node motif has been found in E. coli~\cite{6,9}, yeast~\cite{milo,10}, and other organisms~\cite{11,12,13,14,15,16}. Other motifs have been identified with varieties of memory switches to control and delay actions, among other possible useful biological functions.


FFLs are only one class of network motif among 4 that have been identified~\cite{alonreview}. The simplest kind of motif is the positive and negative autoregulation (with a self-loop as motif), abbreviated NAR and PAR respectively. There are also \textit{Single-input modules} (SIM) and \textit{Dense overlapping regulons} (DOR) or \textit{Multi-input motifs} (MIMs). The SIM is a regulator that regulates a group of genes with no other regulator in between (with the only regulator regulating itself), and is identified with a coordinated expression function of a group of genes whose shared function is capable of generating a temporal expression program and activation order with different thresholds for each regulated gene. The DORs have been identified with a gate-array, carrying out a computation from multiple inputs through multiple outputs.


As a proof-of-principle, we have previously shown that topological properties of complex networks are indeed detected by approximations to $K$~\cite{zenilgraph}. For Kolmogorov complexity, for example, the simplest possible motif is that which connects all nodes bi-directionally, which is consistent with the idea that such a motif would be biologically meaningless (unless assigning weights indicating levels of regulation).

There is a debate, however, as to whether motif analysis is living up to its promise of breaking down complex biological networks in order to understand them in terms of minimal functions carried by these motifs. Indeed it has been suggested that the motif approach to networks has important limits (e.g.~\cite{GRN}).



Even when there are some known facts about how classes of motifs connect to each other, such as FFLs and SIMs integrated into DORs and DORs occurring in a single layer (there is no DOR at the output of another DOR)~\cite{alonreview}, we think the study of motifs and full networks are not alternatives but are complementary approaches, as one cannot reconstruct global patterns from local repetitions, nor local patterns from global properties.


\section{Compressibility of biological networks}

We know that for a random (Erd{\"o}s-R\'enyi) graph $G$, $K(G) \sim |E(G)| = {{n}\choose{2}}$, because edges being placed at random need to be described one by one in order to reconstruct the network from its ``compressed" description without loss of information. Likewise, a complete graph $c$ of size $|V(c)|$ will have a Kolmogorov complexity $K(c)=\log_2 |V(c)|$, because its description is trivial--everything is connected--and only the size of a complete graph is needed in order to reproduce it in full. This means that a statistical (Erd{\"o}s-R\'enyi) graph is also algorithmically (Kolmogorov) random if it is not produced recursively (e.g. produced by an algorithm) but by a random process (most random computer-generated graphs are actually simulations of valid statistically random graphs), while a regular (e.g. a complete) graph has the lowest Kolmogorov complexity. Any other type of graph will lie between these theoretical extremes, so we can expect biological networks to have intermediate values between simplicity and maximum algorithmic randomness, or in other words, we can expect their randomness deficiency to be removed both from simplicity and randomness.

\begin{table}[h] 
\caption{Compression lengths (using Deflate) of a sample of 20
metabolic networks where $|V(G)|$ is the number of vertices of $G$,
$|E(G)|$ the number of edges of $G$, $C(c_{|V(G)|})$ the compressed size of the adjacency matrix of the complete graph with number of nodes equal to $|V(G)|$, $C(G)$ the compressed size of $A(G)$, and $C(r_{|V(G)|})$ the compressed size of the adjacency matrix of a randomized network $r$ built with the same number of vertices as $G$ and also following the degree distribution of $G$.
Networks are sorted by $C(G)$.}\label{table}
\begin{tabular*}{\textwidth}{@{\extracolsep{\fill}}lclclclclclc}
\hline
\textnormal{Network ($G$)} & \textnormal{$|V(G)|$} & \textnormal{$|E(G)|$} &
\textnormal{$C(c_{|V(G)|})$} & \textnormal{$C(G)$} & \textnormal{$C(r_{|V(G)|})$} \\ \hline
\textnormal{Chlamydia Pneumoniae} & 387 & 792 & 0.0020 & 0.032 & 0.90 \\
\textnormal{Mycoplasma Pneumoniae} & 411 & 936 & 0.0018 & 0.034 & 0.89 \\
\textnormal{Chlamydia Trachomatis} & 447 & 941 & 0.0015 & 0.029 & 0.88 \\
\textnormal{Rickettsia Prowazekii} & 456 & 1014 & 0.0014 & 0.030 & 0.89 \\
\textnormal{Mycoplasma Genitalium} & 473 & 1060 & 0.0013 & 0.029 & 0.89 \\
\textnormal{Treponema Pallidum} & 485 & 1117 & 0.0013 & 0.028 & 0.89 \\
\textnormal{Aeropyrum Pernix} & 490 & 1163 & 0.0012 & 0.029 & 0.89 \\
\textnormal{Oryza Sativa} & 665 & 1514 & 0.00068 & 0.022 & 0.92 \\
\textnormal{Arabidopsis Thaliana} & 694 & 1593 & 0.00062 & 0.020 & 0.93 \\
\textnormal{Pyrococcus Furiosus} & 751 & 1768 & 0.00058 & 0.019 & 0.95 \\
\textnormal{Pyrococcus Horikoshii} & 767 & 1796 & 0.00055 & 0.018 & 0.95 \\
\textnormal{Thermotoga Maritima} & 830 & 1980 & 0.00047 & 0.018 & 0.96 \\
\textnormal{Emericella Nidulans} & 916 & 2176 & 0.00039 & 0.016 & 0.98 \\
\textnormal{Chlorobium Tepidum} & 918 & 2159 & 0.00039 & 0.016 & 0.98 \\
\textnormal{Helicobacter Pylori} & 949 & 2325 & 0.00036 & 0.016 & 0.98 \\
\textnormal{Campylobacter Jejuni} & 946 & 2257 & 0.00036 & 0.016 & 0.97 \\
\textnormal{Neisseria Meningitidis} & 981 & 2393 & 0.00034 & 0.015 & 0.99 \\
\textnormal{Porphyromonas Gingivalis} & 1010 & 2348 & 0.00032 & 0.014 & 1.0 \\
\textnormal{Enterococcus Faecalis} & 1004 & 2462 & 0.00032 & 0.016 & 1.0 \\
\textnormal{Streptococcus Pyogenes} & 1051 & 2577 & 0.00030 & 0.014 & 1.0 \\
\hline
\end{tabular*}
\end{table}

As shown in Table.~\ref{table}, Kolmogorov approximation values by incompressibility (denoted by $C$) lie right between the two extreme cases, fully connected ($c$) (or fully disconnected) graphs, because the size of the complete graph $c$ is equal to $G$; and randomized graphs (denoted by $r$) are those constructed by randomly rewiring the network in a way that preserves the original degree distribution of $G$. That biological networks are neither random nor simple captures the fact that they contain non-trivial information as models or descriptions of non-stochastic processes of the living systems they represent. In other words, a biological network $G$ has Kolmogorov complexity $K(G)$ such that $0\sim K(c_{|V(G)|})<K(G)<K(r_{|V(G)|})\sim 1$, where $r_{|V(G)|}$ is a randomized version of $G$ and $c_{|V(G)|}$ the complete graph of size $|V(G)|$. For example, the Streptococcus pyogenes metabolic network with only $\sim 2500$ links reaches greater complexity than the corresponding random graph for the same number of edges, and lies between the simplest cases (disconnected and complete graphs) and random graphs reaching a compression ratio of 1. With only 2\,577 links, however, the complexity of the Streptococcus pyogenes network is far removed from simplicity and is closer to algorithmic randomness (see Table~\ref{table}). In Fig.~\ref{concave}, a simulation of a random $ER$ graph of the size of the Streptococcus pyogenes network was undertaken. It can be seen that lossless compressions follow the entropy curve trend very accurately, as they are actually implementations of variations of an entropic measure, given the way they are designed (based on dictionaries that can only see statistical patterns). The BDM, on the other hand, is truly algorithmic and provides a richer picture of the graph trajectory. 

\section{Robustness of Kolmogorov graph complexity}

Despite the complexity of the calculation of unlabeled complexity $K^\prime$, regular graphs have been shown to have low $K$ and random graphs have been shown to have high $K$ estimations, with graphs with a larger set of automorphisms having lower $K$ than graphs with a smaller set of automorphisms~\cite{zenilgraph}. An important question is how accurate a labeled estimation of $K(G)$ is with respect to the unlabeled $K^\prime(G)$, especially because in the general case the calculation of $K(G)$ is computationally cheap as compared to $K^\prime(G)$, which carries an exponential overhead. However, the difference $|K(G)-K^\prime(G)|$ is bounded by a constant. Indeed, there exists an algorithm $\alpha$ of fixed length size $|\alpha|$ bits such that one can compute all $L(G)$ relabelings of $G$, even if by brute-force, e.g. by producing all the indicated adjacency matrix row and column permutations. Therefore $|K(G)-K(G_L)| < |\alpha|$ for any relabeled graph $G_L$ of $G$, or in other words, $K(G_L) = K^\prime(G) + |\alpha|$, where $|\alpha|$ is independent of $G$. Notice, of course, that here the time complexity of $\alpha$ believed to not be in $\mathbf{P}$ is irrelevant; what is needed for the proof is that it exists and is therefore of finite size. We can therefore safely estimate the unlabeled $K^\prime(G)$ by estimating a labeled $K(G_L)$ as an accurate asymptotic approximation. In fact brute-force is likely the shortest program description to produce all relabelings, and therefore the best choice to minimize $\alpha$.

This result is relevant, first because it means one can accurately estimate $K_L(G)$ through $K(G)$ for any lossless representation of $G$ up to an additive term. One problem is that this does not tell us the rate of convergence of $K(G)$ to $K_L(G)$. But numerical estimations show that the convergence is in practice fast. For example, the median of the BDM estimations of all the isomorphic graphs of the graph in Fig.~\ref{isotests} is 31.7, with a standard deviation of 0.72. However, when generating a graph, the BDM median is 27.26 and the standard deviation 2.93, clearly indicating a statistical difference. But more importantly, the probability of a random graph having a large automorphism group count is low, as shown in~\cite{zenilgraph}, which is consistent with what we would expect of the algorithmic probability of a random graph--a low frequency of production as a result of running a \textit{turmite} Turing machine. And here and in~\cite{zenilgraph} we have also shown that graphs and their dual and cospectral versions have similar Kolmogorov complexity values as calculated by algorithmic probability (BDM), hence indicating that in practice the convergence guaranteed by the result in this section is fast.

\subsection{$K(G)$ is not a graph invariant of $G$}

$K(G)$ may be a computationally cheap to approximate up to a bounded error that vanishes in the size of the graph. However, $K(G)$ does not uniquely determine $G$. Indeed, two non-isomorphic graphs $G$ and $H$ can have $K(G)=K(H)$. In fact the algorithmic Coding theorem gives an estimation of how often this happens, and it is also related to a simple Pigeonhole argument. Indeed, if $G$ or $H$ are algorithmically (Kolmogorov) random graphs, then the probability that $K(G)=K(H)$ grows exponentially. If $G$ and $H$ are complex, then their algorithmic probability $\sim 1/2^K(G)$ and $\sim 1/2^K(H)$ is small and are in the tail of the algorithmic probability (the so called \textit{Universal distribution} or Levin's \textit{semi-measure}) distribution ranging across a very small interval of (maximal) Kolmogorov complexity, hence leading to greater chances of value collisions.

\section{Detection of graph properties}

Properties with threshold values are said to exhibit a ``phase transition'', due to a principle analogous to the one observed in physical systems. One initial question is whether Kolmogorov complexity can detect the first, most basic tradeoff studied in $ER$ networks, that is, a classical graph phase transition of connectivity. 

\begin{figure}[htbp!]
\centering
\includegraphics[width=6cm]{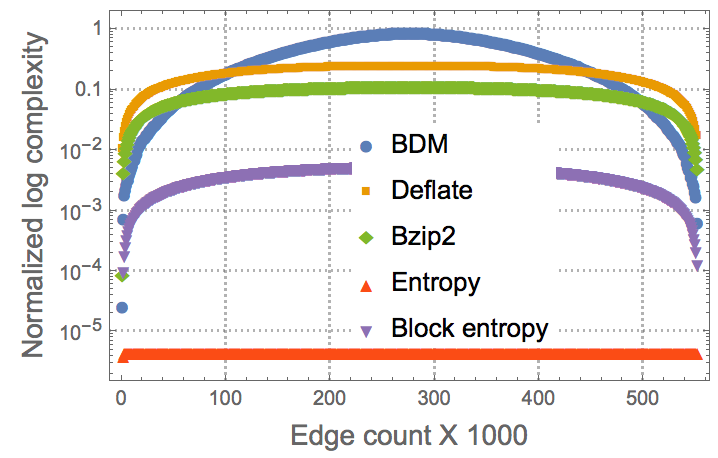} \includegraphics[width=5.7cm]{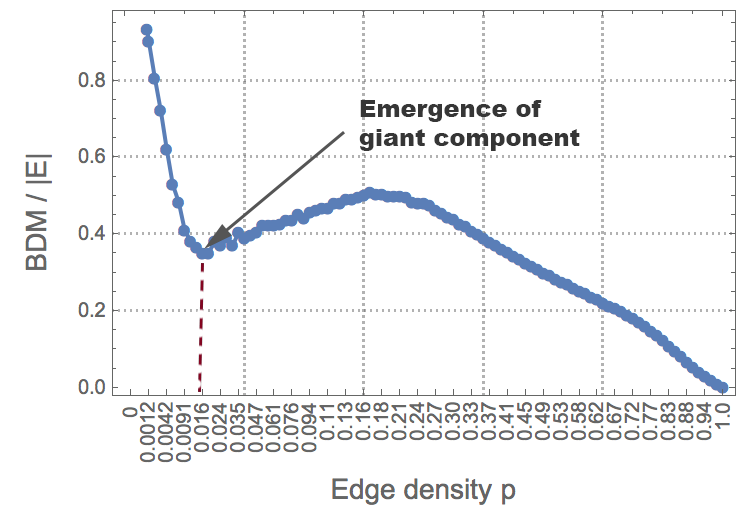}\\
\bigskip
\includegraphics[width=5.9cm]{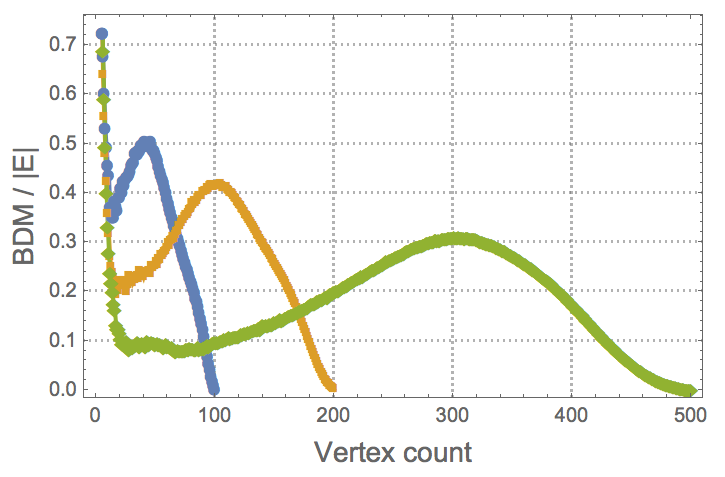} \includegraphics[width=5.7cm]{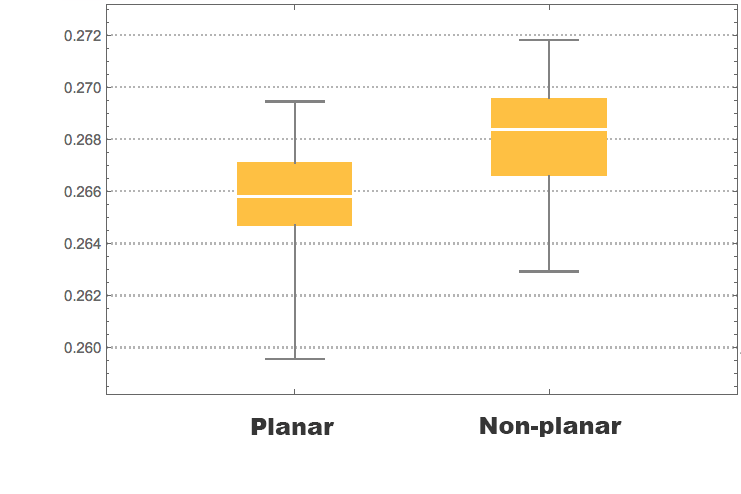}
\caption{
Top left: Curves of complexity of a network of 1051 nodes (the size of the Streptococcus pyogenes metabolic network) when varying the number of edges from 0 to the complete graph with 551\,775 links, normalized by a constant. The maximum complexity is reached when for ${{n}\choose{2}}/2$, i.e. edge density 50\%. Top right: Complexity of an averaged curve of 10 $ER$ graphs of size 100 nodes as a function of edge density normalized by a edge count. The curve is negatively skewed due to the connectivity \textit{phase transition} upon emergence of the ``giant component'' (growing $K$). Bottom left: Complexity curves of three $ER$ graphs (10 replicates each) of growing size (100, 200 and 500) showing that the maximum complexity after the emergence of the giant component asymptotically approaches 50\% edge density and the detection of the giant component corresponds to the theoretical prediction $\log n/n$. Bottom right: Whisker plot of 2\,144 planar and 4\,600 non-planar graphs with significantly different complexity values (for both BDM and compression algorithms).}
\label{concave}
\end{figure}

When the graph remains mostly disconnected (i.e. before the phase transition), its Kolmogorov complexity is close to $K(G)=\log |V|$, but as soon as a ``giant component" emerges at about edge probability $\frac{\log |V(G)|}{V(G)}$ according to the Erd{\"o}s-R\'enyi theorem~\cite{erdos2}, $K(G)$ starts growing driven by $|E|$, as can be seen in Fig.~\ref{concave}. Erd{\"o}s and R\'enyi showed that if $V(G)p \rightarrow c > 1$, where $c$ is a constant, then an $ER$ random graph $G$ with $\binom{n}{2}p$ edges on average would almost certainly have a unique giant component containing a positive fraction of the vertices, and no other component will contain more than $O(log(n))$ vertices, with $V(G)=n$ and $p \sim \frac{\log n}{n}$ a sharp threshold for the connectedness transition of $G(n)$.

Let $\mathbb{P}(x) = a_1x+a_2x^{2}+\ldots a_nx^{n}$ be a polynomial of degree $n>2$ fitting the data points $x_i=K(G(p))$ of a graph $G$ with edge density $p<0.5$ and graph size $|V(G)|=n$. Then the phase transition is found in the local minimum of the non-convex fitted curve given by

$$
\frac{d \mathbb{P}(x)}{d x} = 0\textit{ and } \frac{d^2\mathbb{P}(x)}{d x^2} \geq 0
$$

When undertaking the experiment on a large set of $ER$ graphs, the transition is exactly where the inflection point of the local minimum of a fitted curve of at least degree 3 appears that can be exposed in both log complexity and normalized plots, as seen in Fig.~\ref{concave} (top right and bottom left). This is because $K(g)$ can only grow if non-recursive, i.e. random information is introduced, although what we have in these experiments is recursive pseudo-randomness. Isolated edges remain of low complexity before the emergence of the \textit{giant component}, and isolated graphs would eventually show repetitions if they are not allowed to expand and produce larger components having almost a fourth of the possible number of edges before the network starts decreasing in complexity. This subtle behavior is more difficult to reveal if the complexity is normalized by a constant, as in Fig.~\ref{concave} (top left).

Because the threshold is sharp in the phase transition of the giant component, no changes to either the local or global distribution of 1s in the adjacency matrix or changes in the uniform distribution of the degree sequence are detected by measures such as Shannon entropy or entropy rate. However, for algorithmic complexity the main feature for the description of the graph emerges, and is therefore successfully captured, as seen in Fig~\ref{concave} (top right). One can design a measure~\cite{smalldata} to detect the giant component transition, even an entropic one, by taking as random variable the number of connected links or an equivalent description, but algorithmic complexity does it without changing either the focus of the measure or the description of the object, because it is a robust measure invariant to different object descriptions~\cite{smalldata}. The giant component transition is only a property to illustrate the power of algorithmic information theory, and Fig.~\ref{concave} (top right) shows that there are numerical approaches that conform to theoretical expectations.

There are other properties of graphs related to graph embeddings, having to do with whether the edges of a graph do not cross each other on a plane. Hence in a topological sense it is simpler when the graphs are drawn in a single plane. Interestingly, both BDM and lossless compression consistently assigned lower Kolmogorov complexity to planar graphs, as shown in Fig.~\ref{concave} (bottom right). The 6744 graphs used and classified as planar and non-planar graphs were taken from the repository function \textbf{GraphData[]} available in the Wolfram\textit{Mathematica} software v.10. In~\cite{zenilgraph}, it was also shown that graphs follow the approximations by BDM and lossless compression to the Kolmogorov complexity of their dual graphs, which is as theoretically expected, given that there is an algorithm of fixed size for producing a dual from a graph.

\section{The Kolmogorov complexity of complex networks}
\label{complexnetworkssection}

\begin{figure}[htbp!]
\centering
\includegraphics[width=5.9cm]{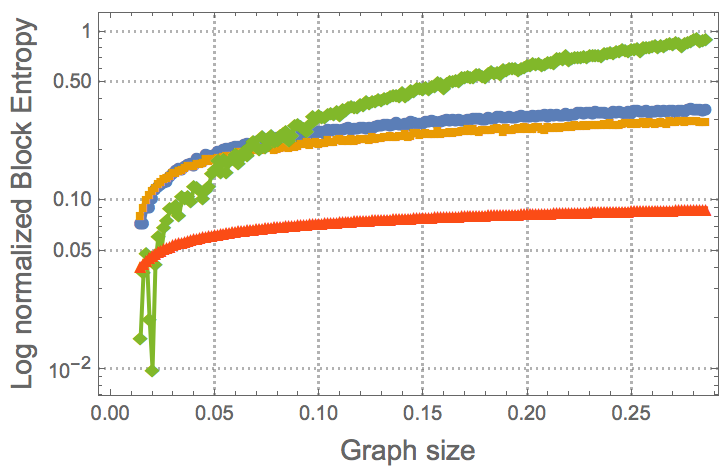} \hspace{.1cm} \includegraphics[width=5.9cm]{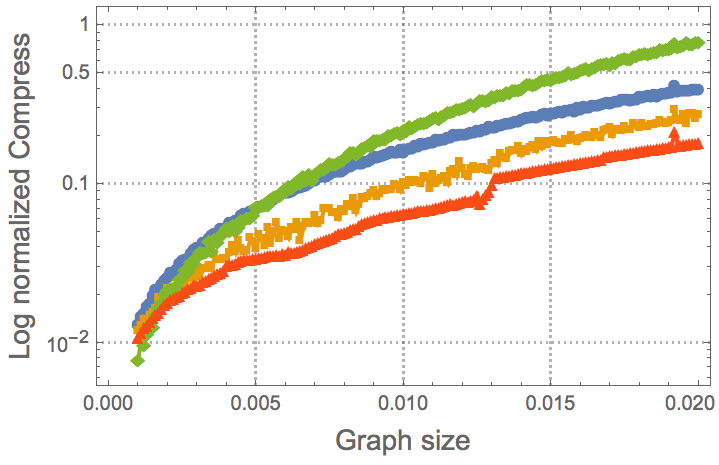}\\
\bigskip
\includegraphics[width=8.2cm]{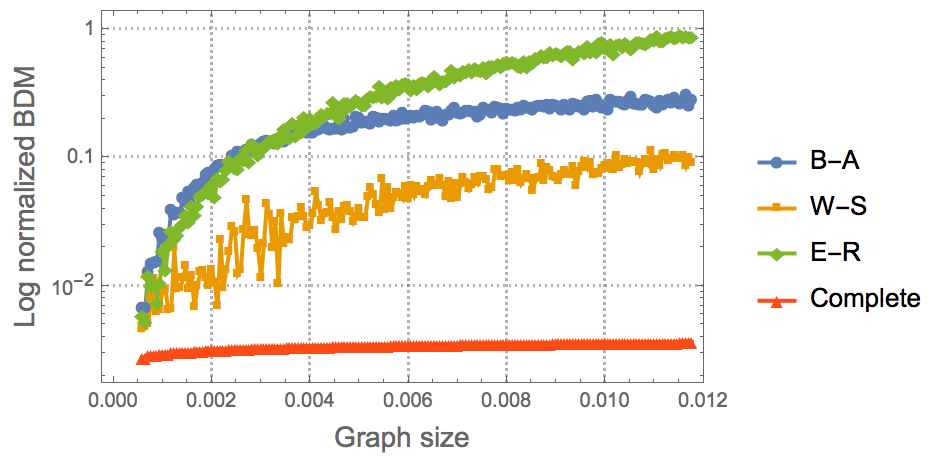}

\caption{Numerical approximations to different network topologies of increasing size, with BDM following the theoretical calculations. For WS, the rewiring probability $p=|V(WS)|/2000$, showing how it moves from complexity values close to the simple (complete) growing graph to the random ER. $ER$ graphs upper bound the complexity of all others, while BA graphs grow slowly by a factor of $\log |E(BA)|$, and the complete graph $c$ is an example of a simple graph (any regular graph) with minimal complexity growing by $\log |V(c)|$. Lossless compression fails to keep regular graphs converging to low complexity as numerically calculated with both Compress and BZip2.}
\label{complexnetworks}
\end{figure}

We are now equipped to calculate the Kolmogorov complexity of a complex network. We have already theoretically calculated and experimentally approximated the Kolmogorov complexity of trivial/simple (denoted here by $S$) and random Erd{\"os}-R\'enyi (ER) graphs. Regular graphs, such as completely disconnected or complete graphs, have Kolmogorov complexity $K(S)= \log |V(S)|$. $ER$ graphs have maximal complexity, so any other complex network is upper bounded by $K(ER)$ graphs. Now the Kolmogorov complexity of a Barab\'asi-Albert (BA) network is low, because it is based on a recursive procedure, but there is an element of randomness accounted for by the attachment probability. If $BA$ is a BA network, then $K(BA)\sim \gamma + \mathcal{O}(|V(BA)|)$, where $\gamma$ is the size of the computer program or Turing machine implementing the preferential attachment algorithm and $\mathcal{O}(|V(BA)|)$ accounts for the attachment probability. However, if $WS$ is a Watts-Strogatz network and $p$ is the edge rewiring probability, if $p \rightarrow 1$, then $K(WS) \rightarrow K(ER)$ and if $p \rightarrow 0$, then $K(WS) \rightarrow K(G)$. We therefore have the following relationships between the most studied graph types and complex networks:

$$
K(S) < K(BA) < K(ER)
$$
$$
K(WS) < K(BA) \textit{ if $p \sim \epsilon$}
$$
$$
K(WS) = K(S) \textit{ if $p = 0$}
$$
$$
K(WS) = K(S) \textit{ if $p = 1$}
$$
$$
K(WS) > K(BA) \textit{ if $p \sim 1-\epsilon$}
$$

\noindent with $\epsilon$ a typical small rewiring probability $< 0.1$. In Table~\ref{table2}, theoretical Kolmogorov complexity values and their numerical approximations are summarized and numerical estimations only are illustrated in Fig.~\ref{complexnetworks} and Fig.~\ref{concave}.

\begin{table}[h] 
\caption{Theoretical calculations of $K$ for different network topologies for $0\leq p \leq 1$. Clearly maximum $K$ is reached for random $ER$ graphs with edge density $p = .25$ for which $K(ER)={{|V(ER)|}\choose{2}}/2$.}\label{table2}
\begin{tabular*}{\textwidth}{@{\extracolsep{\fill}}lclc}
\hline
\textbf{Graph/network} & \textbf{Notation} & \textbf{$K$}\\ \hline
Regular & $S$ & $K(S)= \mathcal{O}(\log |V(S)|)$ \\
Barab\'asi-Albert &$BA$ & $K(BA)= \mathcal{O}(|V(BA)|) +c$\\
Watts-Strogatz &$WS$ & $\lim_{p\rightarrow 0} K(WS) \sim K(S)$\\
&&$\lim_{p\rightarrow 1} K(WS) \sim K(ER)$\\
Random Erd{\"os}-R\'enyi &$ER$ & $K(ER) = \mathcal{O}(\frac{n(n-1)}{16 p|p-1|}) $ \medskip \\
\hline
\end{tabular*}
\end{table}

While $K(ER)$ grows by the number of edges ${{|V(ER)|}\choose{2}}$ given by $p \times n(n-1)/2$ as a function of edge density $p$, $K(ER)$ reaches its maximum at $p=.5$ and then comes back to low $K$ values when approximating the complete graph, because from $p = .25$ on one can start describing the network by the number of missing links rather than the existing ones. Hence the curve for $K(ER)$ is concave, with maximum ${{|V(ER)|}\choose{2}}/2$ (half the possible links) and minimum $K(ER)=\log |V(ER)|$ for $p=0$ and $p=1$. A function approximating the behavior of $ER$ graphs that provide an upper bound of $K$ for any graph, as given in Table~\ref{table2}, can be calculated by interpolating a polynomial of degree two for these maximums and minimums for the two $p$ points which give the quadratic curve $K(ER) = \mathcal{O}(\frac{n(n-1)}{16 p|p-1|})$ as theoretical approximation to $K(ER)$. This would be in agreement with the numerical behavior shown in Figs.~\ref{complexnetworks} and~\ref{concave}, except for the connectivity phase transition of the emergence of the \textit{giant component}, which causes the concave curve to be negatively skewed with a leading tail due to the low growth of $K$ for a mostly disconnected graph before the emergence of the giant component, as the fitted curve in Fig.~\ref{concave} shows.

\subsection{Error estimation of finite approximations}

\begin{figure}[htbp!]
\centering
\includegraphics[width=10cm]{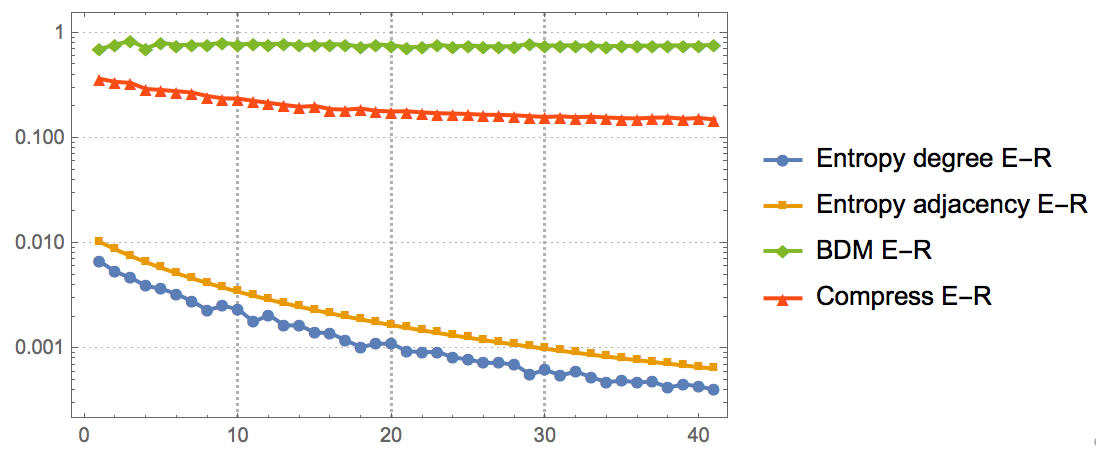}\\
\medskip
\includegraphics[width=9.7cm]{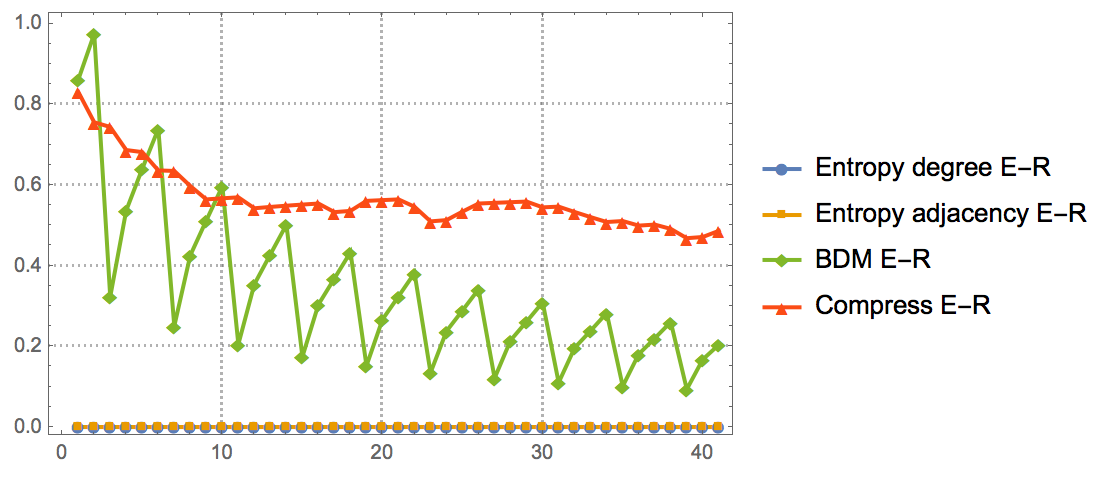}
\caption{Numerical approximations (normalized between 0 and 1) for analytically solvable extreme cases: $ER$ random (edge density 0.5) and complete graphs. Entropy values are exact calculations but BDM and Compress are upper bounds of $K(G)$.}
\label{errortests}
\end{figure}

One important question concerns the error estimation of approximations to the Kolmogorov complexity of a network using both available methods---lossless compression and algorithmic probability. Given that we know the Kolmogorov complexity of extreme graphs (e.g. a complete graph vs. an Erd{\"o}s-R\'enyi graph), Figs.~\ref{errortests} show that the two methods have complementary capabilities when applied to the adjacency matrices or degree sequences of the graphs.

The Kolmogorov complexity of a minimally complex graph (e.g. a complete or empty graph) grows by $K(G)=\log |V(G)|$ because there is no need for a description of the position of any edge (for, e.g. $|E(G)|={{n}\choose{2}}$ or $|E(G)|=0$) and therefore $K(G)$ only grows by the logarithm of the number of vertices $|V(G)|$, while an upper bound of the Kolmogorov complexity of an $ER$ random graph with edge density $0.5$ is $K(G)= |E(G)|$, because one has to specify the end points of every edge.

Fig.~\ref{errortests} shows classical and algorithmic complexity approximations of $ER$ random graphs and complete graphs as calculated by three measures: BDM, lossless compression (Compress) and Entropy (as applied both to the adjacency matrix and to the degree sequences). 

On the one hand BDM and Compress estimate $K(G)$ from above, converging to a value $K(G)/|E(G)| \sim 1$ for $ER$ random graphs and $K(G)/\log|V(G)| \sim 0$ for complete graphs (with BDM converging faster even if it is less stable due to the block size, and for which the error is therefore estimated, allowing correction). 

On the other hand, Entropy as applied to adjacency matrices and degree sequences is similar, and conforms to the expected theoretical values. Entropy decreases for random graphs because $ER$ graphs have the same average node degree, and the frequency of non-zero values in their adjacency matrices is 0.5 from the graph density. 

For complete graphs (with self-loops, hence adjacency matrices with all entries 1) and with the same $V(G)$ edges per node, Entropy is clearly 1 for growing adjacency matrix as well as degree sequences. However, it can be seen that degree entropy and adjacency matrix entropy do not always coincide for random graphs, because of the small variance in node degree, while the adjacency matrix retains the exact edge density 0.5. Unlike theoretical values of Kolmogorov complexity, Entropy is therefore less robust as a measure of the complexity of information content, and while it has the advantage of being computable, its chief advantage is that it is not as powerful as Kolmogorov complexity~\cite{smalldata}.

\subsection{Linear versus algorithmic complexity}

\begin{figure}[htbp!]
\centering
\includegraphics[width=7.5cm]{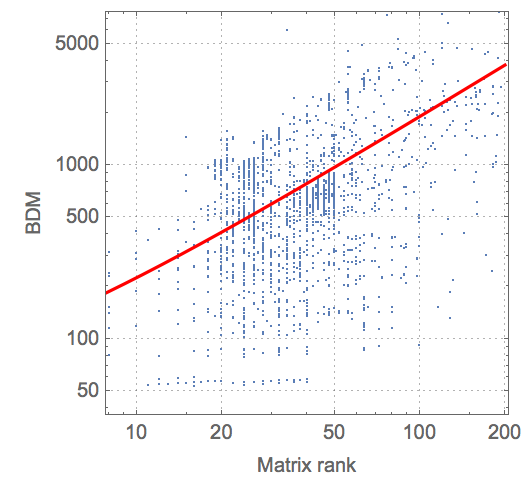}
\caption{Log-log correlation plot from almost 3000 graphs from the GraphData[] repository in Wolfram \textit{Mathematica} v.10 (including graphs) with node count between 20 and 200. confirming the expectation of correlation between numerical approximations to algorithmic complexity and the rank of adjacency matrices of regular graphs with different low algorithmic information content. The linear fitting line (in red) is $37.78 + 18.7 x$ and the Pearson correlation is 0.39 with $p$-value $2.8 \times 10^{-108}$. When normalizing by graph size (node count), the correlation is still strong with a fitting line $1.44 + 19.15 x$, Pearson correlation 0.185 and $p$-value $6.51 \times 10^{-24}$.}
\label{linearcomplexity}
\end{figure}

Based in the ideas of linear complexity described in Section~\ref{classical}, we calculated the matrix rank (the number of linearly independent rows) of a set of regular graphs with different algorithmic complexity (as calculated by BDM) to find the expected correlation as shown in Fig.~\ref{linearcomplexity}.

\section{Algorithmic complexity of synthetic data and artificial networks}
\label{mendes}

Taking advantage of the robustness of Kolmogorov complexity as asymptotic invariant to different descriptions of an object (e.g. a network versus the data from which the network is constructed e.g. gene expression), we can  attempt to find if there is a numerical correspondence of the complexity values of equivalent objects. 

The Mendes dataset~\cite{mendes} is a common and widely used set of artificial gene networks for the objective comparison of network reconstruction analyses. In the database there are files emulating microarrays of gene expression data from interactions determined by particular topological and kinetic properties. These networks are embodied in kinetic models designed to produce synthetic gene expression data used for \textit{in silico} experiments.

The networks consist of three sets of 100 genes, with each set having 200 interactions. They are classified as Erd{\"o}s-R\'enyi random networks (ER), scale-free networks (SF), as defined by Barab\'asi-Albert's preferential attachment algorithm, and small world or Watts-Strogatz networks (WS), depending on which of the three topologies was employed to generate the data and the networks. The Mendes models use a multiplicative Hill kinetics to approximate transcriptional interactions determined by the following differential equation:

$$\frac{d x_i}{dt} = a_i \prod_{j=1}^{N_I} \frac{1K_j^{n_j}}{1K_j^{n_j}+I^{n_j}_j} \prod_{l=1}^{N_A}(1+\frac{A_l^{m_l}}{AK_l^{m_l}+A_l^{m_l}})-b_ix_i$$

\noindent where $x_i$ is the concentration (expression) of the $i$-th gene, $N_I$ and $N_A$ are the number of upstream inhibitors and activators respectively, and their concentrations are $I_j$ and $A_I$. All other parameters are specified and explained in~\cite{mendes}. There is one differential equation for each gene in the network following the behavior of Hill kinetics.

\begin{figure}[htbp!]
\centering
\includegraphics[width=11cm]{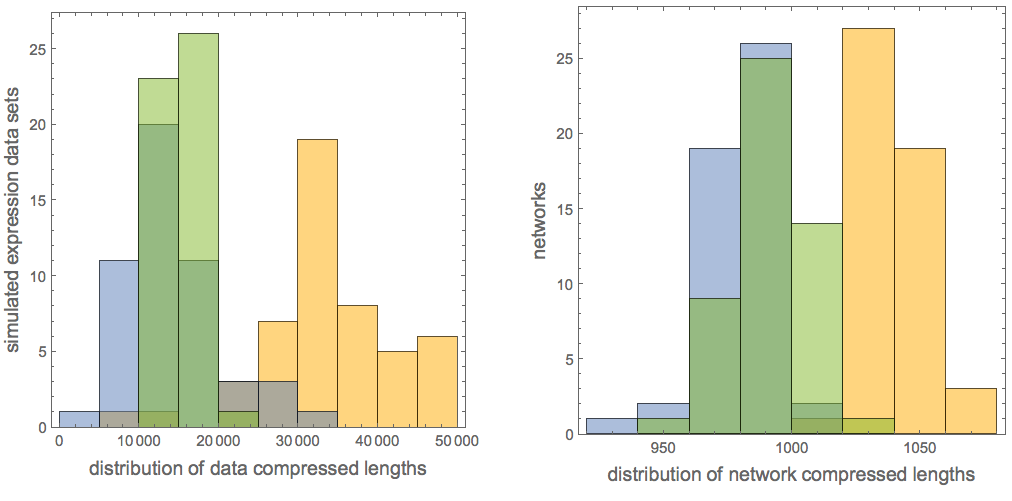} \includegraphics[width=6.5cm]{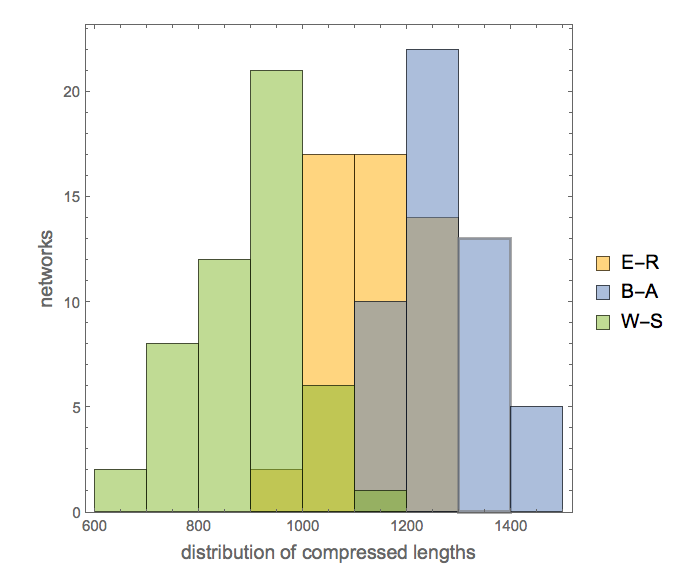}
\caption{Top: Comparison of histograms of compressibility of the simulated gene expression data files and their underlying networks. The data sets and networks were produced by a Hills equation process. Here we show that compression directly on the data and the networks characterizes their topology. Bottom: BDM, however, classifies $ER$ random networks as less complex than scale free networks, which in principle should not be the case, but the range of application of BDM does not cover continuous data, as produced in the artificial gene expression data files from the Mendes database.}
\label{mendesfig}
\end{figure}

The data files in the Mendes database are produced by the Hill Kinetics differential equations for which the networks were designed. Hence a correspondence between the complexity of gene expression data in the files and the networks may be expected. Fig.~\ref{mendesfig} shows how compressibility and BDM can be applied to these objects, namely, data sets containing simulated gene expression (real values) data, and how applying compressibility classifies data and networks in agreement with each other. The expectation would be to see a natural correspondence between the time series' complexity and the networks' complexity. However, the range of application of BDM currently covers only discrete data, and Fig.~\ref{mendesfig} shows that it reverses the $ER$ and the $BA$ networks for this Mendes dataset. This effect can be the result of the ratio of data to noise introduced in the input files for the reconstruction methods in connection to the particular topology of a network, noise which is not present in the gold standard network. Another direction for future research is the extension of BDM to continuous data (by e.g. studying the effects of data discretization, such as binning), which would then allow us to compare the complexity of the simulated gene expression that generates the networks.

The approach would suggest that in the task of network reverse engineering, where only the gene expression data is known, the complexity of the data may provide a hint which may be useful in recovering/reconstructing the topology of the causal network if it cannot even substitute the need of explicit network inference to understand how the components of a complex system like a living cell interact and regulate each other and how elements influence each other.

\section{Conclusions}

We have surveyed and introduced several concepts and methods at the intersection of network biology, graph theory, complex networks and information theory. We have applied and compared techniques, from Shannon's entropy to Kolmogorov complexity (by way of algorithmic probability and lossless compressibility), that can be used to analyze various aspects and properties of biological networks, techniques serviceable at different scales and for different purposes.

We have aimed for an encompassing information-theoretic study of biological networks at different scales and for a potentially fruitful interaction between algorithmic information theory and systems biology. 

We have introduced theoretical and numerical results to be further explored and exploited. We have applied the methods to various networks, and in particular to artificial gene expression data files and their associated networks with different topologies.

The process of data discretization of continuous sources is a topic unto itself. Traditionally, it has been done by binning methods or density estimations. A survey of these techniques is offered in~\cite{kotsiantis}. But in the application of information-theoretic measures to sequences of discretized real valued data, it is important to set values to the same digit precision. In the Mendes database, for example, discretization is achieved by simply truncating the values of the evaluations of the Hill equations.

Other important matters to investigate further are the limitations of the measurement process and the sources of noise. Noise is part and parcel of naturally occurring processes, arising either from the measurements themselves or from interactions with other systems. Noise is usually expected to be random (but can be recursive or non-recursive), and it would look random for information-theoretic measures, as it would be alien to the mechanism generating the process of interest that's being measured. However, the nature of this randomness can be an interlaced recursive signal or a disconnected one, and it will ultimately have an impact on the approximations of complexity. Differences in the complexity of different descriptions, as coming, e.g., from the simulated gene expression files in the Mendes network vs. the generated networks themselves, can therefore be due to these factors. For example, truncation for discretization of a continuous process can be both a source of apparent noise or a fundamental loss of important information. In the former case, the Kolmogorov complexity of one system relative to the other would increase, and in the latter case it would decrease, thereby in principle empowering the quantification of these phenomena.

\end{document}